\documentclass[aps,prl,twocolumn,superscriptaddress]{revtex4-2}
\pdfoutput=1

\usepackage[utf8]{inputenc}
\usepackage[T1]{fontenc}

\usepackage{amsfonts}
\usepackage{amsmath}
\usepackage{amsthm}
\usepackage{amssymb}
\usepackage{stmaryrd}
\usepackage{amscd}
\usepackage{eucal}
\usepackage{dsfont}
\usepackage{bm}
\usepackage{graphicx}

\usepackage{hyperref}
\usepackage{booktabs}
\usepackage{microtype}
\usepackage[title]{appendix}

\input{macro.lib}

\allowdisplaybreaks

\begin{document}

\title{Ballistic particle transport and Drude weight in gases}
\author{Frank G\"ohmann} 
\affiliation{Fakultät für Mathematik und Naturwissenschaften, Bergische
Universität Wuppertal, 42097 Wuppertal, Germany}
\author{Andreas Kl\"umper} 
\affiliation{Fakultät für Mathematik und Naturwissenschaften, Bergische
Universität Wuppertal, 42097 Wuppertal, Germany}
\author{Karol K. Kozlowski}
\affiliation{Univ Lyon, ENS de Lyon, Univ Claude Bernard, CNRS,
Laboratoire de Physique, F-69342 Lyon, France}

\begin{abstract}
Owing to the fact that the particle current operator in
non-relativistic gases is proportional to the total momentum
operator, the particle transport in such systems is always
ballistic and fully characterized by a Drude weight $\D$.
The Drude weight can be calculated within linear response
theory. It is given by the formula $\D = 2 \p D$, where $D$
is the density of the gas. This holds in any dimension and
for every equilibrium ensemble, in particular for
generalized Gibbs ensembles that describe possible equilibrium
states of isolated integrable quantum systems. In the
canonical ensemble case, the Drude weight can
be equivalently obtained from a generalized susceptibility
related to the fluctuations of the conserved particle current.
Such susceptibility can be rigorously calculated for the
integrable Lieb-Liniger Bose gas in any generalized Gibbs
ensemble using a generalized Yang-Yang thermodynamic
formalism. The resulting expression agrees with a prediction
made within the context of generalized hydrodynamics. It
also allows us to see explicitly that, within truly
generalized Gibbs ensembles, the conductivity related with the
particle current is \emph{not} determined by the corresponding
current-current auto-correlation function.
\end{abstract}

\maketitle

{\bf Introduction.}
In Galilei invariant many-particle systems the particle current
$\Jv$ is proportional to the total momentum $\Pv$ \cite{Rosch06}.
This rather obvious fact renders the particle transport properties
of gases as calculated within linear response theory \cite{Kubo57}
much simpler than those of crystals. Unlike the particle current
in crystalline solids, in which Umklapp scattering caused by
the broken translation invariance induces a non-trivial
optical conductivity \cite{Rosch06}, the particle current
in gases is always ballistic. This must have been clear to
the founders of linear response theory \cite{Kubo57},
but we could not find it spelled out in any paper older than
\cite{Rosch06}. Most likely in the old days there was simply
little interest in the conductivity connected with the particle
current in gases.

This has changed within the last twenty years when more
exotic ballistic transport phenomena came into the focus
of theoretical and experimental studies, in particular, in
the context of integrable quantum systems in one spatial
dimension \cite{BHKPSZ21}. In integrable quantum spin chains
the energy transport can be purely ballistic, meaning that
the thermal conductivity is fully determined by a thermal
Drude weight. This theoretical result \cite{KlSa02,SaKl03}
is reflected in the highly anisotropic thermal conductivity
of crystals with quasi-one dimensional magnetic substructures
\cite{Hessetal03,Hlubeketal10}.

In \cite{KlSa02} it was realized that, if the current
connected with a conserved charge is itself conserved, then
the corresponding Drude weight can be calculated as a second
derivative of a thermodynamic potential pertaining to 
a generalized Gibbs ensemble (although this term was not in
use by that time). The Drude weight is then interpreted
as a generalized susceptibility that quantifies the
fluctuations of the conserved current and can be calculated
by finite-temperature Bethe Ansatz techniques. The same
strategy can be pursued to calculate the Drude weight
pertaining to the particle current of the the Lieb-Liniger
Bose gas.
 
Integrable quantum systems have an extensive number of
local conserved charges. In general, the corresponding global
current operators do \emph{not} commute with the Hamiltonian.
The local current-current correlation functions are therefore
generally not purely ballistic, but may have a ballistic component
quantified by a Drude weight. In recent years a generalized
hydrodynamics of integrable systems has been developed
\cite{BCDF16,CDY16} that is expected to describe the dynamics
of the local conserved charges and currents as observed in
multi-point correlation functions asymptotically on large
spatio-temporal scales. Drude weights, in particular,
referring to spatially homogeneous situations in the
infinite-time limit are expected to come out exactly.

\enlargethispage{15ex}

A full matrix of Drude weights, defined in terms of
current-current correlation functions (see (\ref{DoSp}) below)
and supposed to connect any conserved current of the Bose
gas with any of the possible driving forces, has been
obtained by Doyon and Spohn in \cite{DoSp17} on the basis
of generalized hydrodynamics. We shall see below that in
the thermal equilibrium case their expression for the
particle-current Drude weight, which is just a single
element of this infinite matrix, agrees with the exact Drude
weight that we obtain from a reasoning along the lines of
\cite{KlSa02}. Beyond standard thermal equilibrium their
expression is still exact, but differs from the Drude weight
dictated by linear response.

{\bf Particle transport in gases.}
Consider a mono-atomic gas in $d$ dimensional space consisting
of $N$ non-relativistic particles, with canonical positions and
momenta $\xv_j$ and $\pv_k$, that interact pairwise with
potential $U(\xv)$. For later convenience we set the particle
mass equal to $m =1/2$. Then the dynamics of the gas is driven
by the Hamiltonian
\begin{equation} \label{gasham}
     H = \sum_{j=1}^N \|\pv_j\|^2 + \sum_{1 \le j < k \le N} U(\xv_j - \xv_k) \epp
\end{equation}
Due to the fact that $\dot \xv_j = \i [H, \xv_j] = 2 \pv_j$,
the center of mass momentum $\Pv = \sum_{j=1}^N \pv_j$ and the
particle current $J = \sum_{j=1}^N \dot \xv_j$ in such a gas are
proportional to one another,
\begin{equation} \label{jp}
     \Jv = 2 \Pv \epp
\end{equation}
Since the Hamiltonian (\ref{gasham}) is translation invariant,
it follows that the particle current is conserved
\begin{equation} \label{jconserved}
     [H, \Jv] = 0 \epp
\end{equation}
This is a unique property of gases which drastically simplifies,
as compared to a crystalline solid, the calculation of the
linear response of the gas to a uniform force field.

A uniform force field $\Fv (t)$ acts identically on all particles,
and therefore affects only the center of mass $\Xv/N$ of the gas,
where $\Xv = \sum_{j=1}^N \xv_j$. Within linear response theory
\cite{SupplGKK25} the complex conductivity tensor is given by
the Kubo formula
\begin{equation} \label{conduct}
     \Si^\a_\be (\om) = \int_0^\infty \rd t \: \re^{- \i \om t}
                        \PH_{J^\a, X_\be} (t)
\end{equation}
where $\PH_{J^\a, X_\be}$ is the response function that describes
the reaction of the current component $J^\a$ to the force 
field component $F^\be$ coupled to the gas through operator $X_\be$.

For the response function of the gas in thermal equilibrium
with a heat bath of temperature $T$ we have the two equivalent
expressions \cite{SupplGKK25}
\begin{subequations}
\begin{align} \label{resp1}
     \PH_{J^\a, X_\be} (t)
        & = - \i \<[X_\be, J^\a (t)]\>_T \\
        & = \int_0^\frac{1}{T} \rd \la \: \<J_\be (- \i \la) J^\a (t)\>_T
	    \label{resp2} \epc
\end{align}
\end{subequations}
where the brackets $\< \cdot \>_T$ denote the thermal average.
If $\Jv = 2 \Pv$, as in our case, then $\PH_{J^\a, X_\be}$ is
time independent and both formulae are easily evaluated. From
(\ref{resp1}) we infer that
\begin{equation} \label{evalresp1}
     \PH_{J^\a, X_\be} = - 2 \i \<[X_\be, P^\a]\>_T = 2 N \de^\a_\be \epc
\end{equation}
while (\ref{resp2}), implies that
\begin{equation}
     \PH_{J^\a, X_\be} = \frac{4}{T} \<P_\be P^\a\>_T
                      = \frac{4}{dT} \<P_\g P^\g\>_T \de^\a_\be \epp
\end{equation}

For a time independent response function the integral in
(\ref{conduct}) trivializes, and
\begin{equation}
     \Si^\a_\be (\om) = \PH_{J^\a, X_\be}
        \biggl(\p \de(\om) - {\cal P} \frac{\i}{\om}\biggr) \epc
\end{equation}
where ${\cal P}$ denotes the principal value. It follows that
\begin{equation} \label{ballsig}
     \Re \Si^\a_\be (\om) = \p \PH_{J^\a, X_\be} \de(\om) \epp
\end{equation}

As we knew beforehand and see from (\ref{resp1}), the conductivity
tensor of gases is isotropic. On physical grounds it must be
extensive. A material specific quantity, the specific conductivity
$\s$, is therefore obtained by taking the trace, dividing by the
volume $V$ times the space dimension $d$, and performing the
thermodynamic limit,
\begin{equation}
     \s (\om) = \lim_{V \rightarrow \infty} \frac{\tr \Si (\om)}{Vd} \epp
\end{equation}
Using (\ref{ballsig}) we see that
\begin{equation}
     \Re \s (\om) = \D \, \de(\om) \epc
\end{equation}
wherein
\begin{subequations}
\label{drudeeq}
\begin{align} \label{eccedw}
     \D & = \lim_{V \rightarrow \infty} \frac{2 \p N}{V} = 2 \p D \\
        & = \lim_{V \rightarrow \infty} \frac{4 \p \<P_\g P^\g\>_T}{V T d}
	\label{eccedweq}
\end{align}
\end{subequations}
is called the Drude weight.

We interpret these findings as follows. Within the linear response
regime the transport of particles in gases is isotropic and
purely ballistic. The term `purely ballistic' refers to the delta
function in frequency. Its meaning becomes clearer if we
consider the corresponding linear relation
in the time domain \cite{SupplGKK25},
\begin{equation} \label{experimentaldw}
     \lim_{V \rightarrow \infty} \frac{\de \<\Jv\>_T (t)}{V}
        = \frac{\D}{\p} \int_{- \infty}^t \rd t' \: \Fv(t') \epp
\end{equation}
The current follows immediately the time-integrated force.
If $\Fv (t)$ has compact support $\de t$, we may interpret
the integral on the right hand side as $|\de t| \<\Fv\>_{\de t}$,
where the brackets denote the time average. This has been
used in the recent experimental work \cite{STBCJSM24} in
order to measure the Drude weight as the ratio of the
current divided by the time-integrated force. The experimental
data seem well compatible with (\ref{eccedw}) (cf.\ Fig.~4a of
\cite{STBCJSM24}).

Equation (\ref{drudeeq}) expresses the Drude weight in two
rather different ways, on the one hand as proportional to
the density of the gas, on the other hand as proportional to
a susceptibility related to the conserved total momentum.
The equality of the two expressions implies that the square
of the fluctuations of the total momentum in a gas is
\begin{equation} \label{freecm}
     \<P_\g P^\g\>_T = \frac{d}{2} T N \epp
\end{equation}

The first formula, $\D = 2 \p D$, had appeared in \cite{VKM15},
where the authors noticed that it is a consequence of
the f-sum rule \cite{Mahan00}, $\int \rd \om \: \s (\om) =
2 \p D$, in a system with conserved total particle current.
Although it is quite generally valid for any system with
a Galilei invariant Hamiltonian and kinetic energy of the
form as in (\ref{gasham}), and not just for mono-atomic gases,
it seems it is not well-known. Interestingly, the same formula
is obtained from the classical Drude formula for the conductivity 
\begin{equation}
     \s_\tau (\om) = \frac{2 \i D}{\om + \i/\tau} \epc
\end{equation}
which is based on a simple mechanical model of particles
subject to an effective friction quantified by the
collision time $\tau$ (cf.~\cite{SupplGKK25}
), in the `dissipationless limit' $\tau \rightarrow + \infty$.

{\bf Beyond thermal expectation values.}
The linear response formalism is very robust. Equations
(\ref{conduct}), (\ref{resp1}) remain valid, if the thermal
expectation value $\< \cdot \>_T$ is replaced by any other
ensemble average $\< \cdot \> = \tr\{\r_0 \cdot\}$ in which 
$\r_0$ is a stationary density matrix, $[H, \r_0] = 0$
(cf.~\cite{SupplGKK25}).
Equation (\ref{evalresp1}) remains
valid, as it is entirely independent of the density matrix.
This means that (\ref{eccedw}) holds for any stationary
density matrix in any dimension, in particular, for generalized
Gibbs ensembles \cite{RDYO07,ViRi16} that describe non-thermal
equilibria of integrable one-dimensional systems.

The same is not true for (\ref{eccedweq}) as the equivalence
of (\ref{eccedw}) and (\ref{eccedweq}) in the thermal case
relies on the fact that the canonical density matrix
$\re^{- \frac{H}{T}}/Z$ is basically the time
evolution operator for imaginary times (cf.\ \cite{SupplGKK25}),
which, in general, does not hold anymore for
generalized Gibbs ensembles. However, in the integrable
case (\ref{eccedweq}) can still be calculated as a generalized
susceptibility by Bethe Ansatz techniques. We shall illustrate
this with the example of the Bose gas with contact interactions
in one dimension \cite{LiLi63}, which is one of the simplest
integrable models. It corresponds to the Hamiltonian
(\ref{gasham}) for $d=1$ and with interaction potential $U(x) =
c \, \de(x)$, $c > 0$. We shall also relate our calculation
with work by Doyon and Spohn \cite{DoSp17} who consider
Drude weights of the Bose gas within a generalized hydrodynamic
approach.

Being integrable the model has a sequence of mutually commuting
conserved quantities $I_j$, $j \in {\mathbb N}$, such that
$I_0 = \hat N$ is the particle number operator, $I_1 = P$,
and $I_2 = H$. We consider generalized Gibbs ensembles with
partition functions
\begin{equation} \label{gengips}
     Z_L (T; \muv) = \tr \bigl\{\re^{- \frac{1}{T} \sum_{j=0}^n \m_j I_j}\bigr\} \epc
\end{equation}
where $L$ denotes the length of the system. $\muv$ is a vector
of higher chemical potentials, $n \ge 2$ is assumed to be even,
$\m_0 = - \m$, $\m_1 =  -\g$, $\m_2 = 1$, and $\m_n > 0$.

With $Z_L$ we associate the generalized Gibbs potential
\begin{equation}
     \PH (T; \muv) = - \lim_{L \rightarrow \infty} \frac{T}{L} \ln Z_L (T; \muv) \epp
\end{equation}
A susceptibility pertaining to the total momentum $P$ is
\begin{equation} \label{phigammatwo}
     \chi_P = - 4 \p \6_\g^2 \PH
            = \frac{4 \p}{T} \lim_{L \rightarrow \infty}
	      \frac{\<P^2\> - \<P\>^2}{L} \epc
\end{equation}
where the generalized Gibbs ensemble average is denoted by
$\<\cdot\>$.

For the Bose gas the potential $\PH$ is accessible through the
Yang-Yang thermodynamic formalism \cite{YaYa69}. It represents
$\PH$ as
\begin{equation} \label{phiyy}
     \PH (T; \muv) = - \frac{T}{2 \pi} \int_{- \infty}^\infty \rd q \:
                         \ln\bigl(1 + \re^{- \frac{\e(q|T; \muv)}{T}}\bigr) \epc
\end{equation}
where $\e$ satisfies the (generalized \cite{MoCa12}) Yang-Yang
equation
\begin{multline} \label{yaya}
     \e (k|T;\muv) = \\
        \sum_{j=0}^n \m_j k^j 
	   - T \int_{- \infty}^\infty \rd q \: K(k - q)
	       \ln\bigl(1 + \re^{- \frac{\e(q|T,\muv)}{T}}\bigr) \epp
\end{multline}
Here $K(k) = \frac{c}{\p (c^2 + k^2)}$ is the kernel function
associated with the Bose gas and $c$ is the strength of the
potential introduced above.

The thermal case corresponds to $n = 2$, $\g = 0$. In that
case $\chi_P$ can be calculated from (\ref{phigammatwo}) by
first taking the derivative and then setting $\g = 0$; and
$\<P\> = 0$ on the right hand side of (\ref{phigammatwo}),
since $\hat N$ and $H$ both have even parity. Moreover, for
$n=2$, the generalized Gibbs potential with $\g$ included
can be related to the usual grand canonical potential by a
shift of the chemical potential: The uniqueness of the
solution of the Yang-Yang equation \cite{Kozlowski14}
implies that
\begin{equation}
     \e(k + \tst{\frac{\g}{2}}|T; - \m, - \g, 1) = \e(k|T; - \m - \tst{\frac{\g^2}{4}}, 0, 1)
\end{equation}
and therefore
\begin{equation}
     \PH(T; - \m, - \g, 1) = \PH(T; - \m - \tst{\frac{\g^2}{4}}, 0, 1) \epp
\end{equation}
It follows that
\begin{multline}
     \chi_P|_{\g = 0}
        = \frac{4 \p}{T} \lim_{L \rightarrow \infty} \frac{\<P^2\>_T}{L} \\
        = - 2 \p \6_\m \PH(T; - \m, 0, 1) = 2 \p D \epc
\end{multline}
and we have re-established (\ref{drudeeq}) by a direct calculation.

{\bf Comparison with a result obtained within generalized
hydrodynamics.}
In their work \cite{DoSp17} Doyon and Spohn consider the
correlation functions
\begin{equation} \label{DoSp}
     D_{m, n} =
     \lim_{t \rightarrow \infty} \lim_{L \rightarrow \infty}
        \frac{\<J_n (t) J_m\> - \<J_n\>\<J_m\>}{L} \epc
\end{equation}
where the $J_n$ are the global currents associated with the
conserved quantities $I_n$ (see equation (1.3) of their paper).
They call the whole semi-infinite matrix with indices $m, n
\in {\mathbb N}$ defined by (\ref{DoSp}) `the Drude weight'.
For $m = n = 0$ we have $J_0 = J = 2P$.

Comparing with (\ref{phigammatwo}) we see that $D_{0, 0} =
T \chi_P/\p = - 4 T \6_\g^2 \PH$ which can be calculated
from (\ref{phiyy}), (\ref{yaya}). First of all, (\ref{phiyy})
implies that
\begin{equation} \label{derphione}
     \6_\g^2 \PH =
        \frac{1}{2 \p} \int_{- \infty}^\infty \rd q \: \Biggl\{
	   \frac{\6_\g^2 \e (q)}{1 + \re^\frac{\e(q)}{T}} -
	   \frac{(\6_\g \e(q))^2 \re^\frac{\e(q)}{T}}
	        {T \bigl(1 + \re^\frac{\e(q)}{T}\bigr)^2} \Biggr\} \epc
\end{equation}
where we have suppressed the dependence of the various functions on
the Lagrange parameters. Taking derivatives of the generalized
Yang-Yang equation (\ref{yaya}) we obtain
\begin{subequations}
\begin{align} \label{lineh}
     \6_\m \e (k) & =
        - 1 + \int_{- \infty}^\infty \rd q \: K(k - q)
	         \frac{\6_\m \e(q)}{1 + \re^\frac{\e(q)}{T}} \epc \\
     \6_\g^2 \e (k) & =
        - \int_{- \infty}^\infty \rd q \: K(k - q)
	      \frac{(\6_\g \e(q))^2 \re^\frac{\e(q)}{T}}
	           {T \bigl(1 + \re^\frac{\e(q)}{T}\bigr)^2}
		   \notag \\ & \mspace{90.mu}
        + \int_{- \infty}^\infty \rd q \: K(k - q)
	     \frac{\6_\g^2 \e(q)}{1 + \re^\frac{\e(q)}{T}} \epp
\end{align}
\end{subequations}
Applying the dressed function formula to the latter two
equations, interchanging the two resulting integrations and
using once more (\ref{lineh}) we obtain the identity
\begin{equation}
     \int_{- \infty}^\infty \rd q \:
        \frac{\6_\g^2 \e(q)}{1 + \re^\frac{\e(q)}{T}} =
           \int_{- \infty}^\infty \rd q \:
	      \frac{(\6_\g \e(q))^2 \re^\frac{\e(q)}{T}}
	           {T \bigl(1 + \re^\frac{\e(q)}{T}\bigr)^2}
		   \bigl(\6_\m \e(q) + 1\bigr) \epp
\end{equation}
Inserting it into (\ref{derphione}), multiplying by $- 4 T$
and substituting as well the equilibrium root density function
\begin{equation}
     \r (k) = - \frac{1}{2 \p} \frac{\6_\m \e(k)}{1 + \re^\frac{\e(k)}{T}}
\end{equation}
we end up with
\begin{equation}
     - 4 T \6_\g^2 \PH = \int_{- \infty}^\infty \rd k \:
	\frac{\r (k) \bigl(2 \6_\g \e (k)\bigr)^2}{1 + \re^{- \frac{\e(k)}{T}}} \epp
\end{equation}

This can now be compared with the formula obtained by Doyon
and Spohn. Inserting the definition \cite{DoSp17}
\begin{equation}\label{ggevelocity}
     v^{\rm eff} (k) = \frac{2 \6_\g \e(k)}{\6_\m \e(k)} \epc
\end{equation}
we see that
\begin{equation} \label{ggedrude}
     - 4 T \6_\g^2 \PH = \int_{- \infty}^\infty \rd k \:
	\frac{\r (k) \bigl(v^{\rm eff} (k)\bigr)^2 \bigl(\6_\m \e(k))^2}
	     {1 + \re^{- \frac{\e(k)}{T}}} \epc
\end{equation}
in full agreement with equation (1.3) of \cite{DoSp17} for all
generalized Gibbs ensembles, however, only for $n=2$, $\g = 0$
the quantity gives the Drude weight.

{\bf Summary and discussion.}
We have shown (or rather recalled) that particle transport in
atomic gases is always purely ballistic and is characterized
by a Drude weight $\D$. Within linear response theory this
Drude weight is temperature independent and satisfies the
universal formula $\D = 2 \p D$, where $D$ is the density of
the gas. This formula holds for any Galilei invariant system
with non-relativistic kinetic energy as in (\ref{gasham})
and for any equilibrium ensemble. In particular, it is also
valid for 1d integrable systems described by generalized Gibbs
ensembles.

In thermal equilibrium the Drude weight is proportional to
a susceptibility (\ref{eccedweq}) connected with the conserved
particle current which can be calculated from a generalized
Gibbs ensemble that includes the total momentum in the set of
conserved quantities. For the Lieb-Liniger Bose gas
this quantity can be calculated explicitly in thermal equilibrium
and for generalized Gibbs ensembles. In the former case the
explicit calculation confirms the general formula (\ref{drudeeq}),
in the latter case it provides a rigorous justification of a
formula obtained by Doyon and Spohn within the context of
generalized hydrodynamics. We would like the reader to keep
in mind, however, that in case of an equilibrium state described
by a generalized Gibbs ensemble their `Drude weight' is
generally not the factor $\D$ in (\ref{experimentaldw}) that
has been measured in a recent cold-atom experiment
\cite{STBCJSM24}. The experiment measures $\D = 2 \p D$,
in accordance with our general argument, which also implies
that it is impossible to infer from the experimental data
in \cite{STBCJSM24} if the system is `pre-thermalized' to a
generalized Gibbs ensemble or has thermally equilibrated
with its environment. In Table~\ref{Tab:nonequiv} we illustrate
the general inequivalence of (\ref{eccedw}) and (\ref{eccedweq})
by numerical examples based on the generalized Yang-Yang
formalism (\ref{phiyy}), (\ref{yaya}).

\begin{table}
\begin{center}
\setlength{\tabcolsep}{.5em}
\begin{tabular}{ccc}
\toprule
    $ $ & $\m = 1 $ & $ \m = 2$ \\ \midrule[1pt]
    $T = 1$ & $0.99997$ & $0.99993$ \\
    $T = 2$ & $0.99994$ & $0.99990$ \\
\bottomrule
\end{tabular}
\hfill
\begin{tabular}{ccc}
\toprule
    $ $ & $\m = 1 $ & $ \m = 2$ \\ \midrule[1pt]
    $T = 1$ & $0.35685$ & $0.29298$ \\
    $T = 2$ & $0.29539$ & $0.26731$ \\
\bottomrule
\end{tabular}
\caption{\label{Tab:nonequiv}
Numerical value of $2 \<P^2\>/TN$ as obtained
from (\ref{phigammatwo})-(\ref{yaya}) for different values
of temperature $T$ and chemical potential $\m$ for $c = 1$.
Deviations from $1$ show that the last equation in
(\ref{drudeeq}) is violated (see (\ref{freecm})). Left
table: thermal case, $n = 2$, $\g = 0$ in (\ref{gengips}).
Right table: a generalized Gibbs case, $n = 4$, $\g = 0$,
$\m_4 = 1$ in (\ref{gengips}). The values in the left table
are equal to $1$ within our numerical accuracy, the values
in the right table clearly differ from $1$. The deviation
from $1$ increases as the influence of the higher conserved
charges becomes stronger, i.e., at higher $T$ and larger $\mu$,
respectively.}

\label{tab_velocity}
\end{center}
\end{table}

We would like to emphasize once again that integrability
is not necessary for purely ballistic transport. It is
sufficient to have a single global current that commutes
with the Hamiltonian. We find it interesting that for gases
the particle transport is purely ballistic, while for
integrable spin chains with $R$-matrices of difference form
it is the energy transport that is always purely ballistic.

We are aware of a few more connections to previous works
that we would like to mention. For 1d gases in thermal 
equilibrium (but not in arbitrary generalized Gibbs ensembles)
equation (\ref{eccedw}) can be interpreted as $\D = 2 k_F$,
where $k_F$ is the Fermi momentum. This form of (\ref{eccedw})
was derived in several places \cite{Giamarchi03,Bortz06,BSAS24}
for $T = 0$ by means of Bosonization. In particular, the
particle Drude weight of the $q$-Boson chain, an integrable
discretization of Lieb-Liniger model, was studied in
\cite{Bortz06}, where (\ref{eccedw}) was obtained in the
continuum limit at $T = 0$. A technique to calculate the
spin Drude weight of the XXZ quantum spin chain as a thermal
average over the Kohn formula \cite{Kohn64} was devised in
\cite{BFKS05}. As is explained in the Supplemental material
\cite{SupplGKK25}, that also contains the additional references 
\cite{AsMe76, FuKa98,Kubo57,LBPG25pp,Zotos99}, it can be applied
to the Bose gas and reproduces the results laid out above.

{\bf Acknowledgment.} The authors would like to thank
Imke Schneider and Sergei Rutkevich for helpful discussions.
References \cite{Bortz06,STBCJSM24} were brought to their
attention by Imke Schneider and reference \cite{VKM15}
by one of the unknown referees, whom the authors wish to express
their gratitude. FG acknowledges financial support by the
Deutsche Forschungsgemeinschaft through an ANR-DFG grant number
Go 825/13-1. KKK is supported by CNRS-IEA ``Corr\'elatieurs
dynamiques de la cha\^ine XXZ en et hors \'equilibre thermal'',
by an ANR-DFG grant number ANR-24-CE92-0033 and by the
ERC Project LDRAM: ERC-2019-ADG, Project 884584.


\providecommand{\bysame}{\leavevmode\hbox to3em{\hrulefill}\thinspace}
\providecommand{\MR}{\relax\ifhmode\unskip\space\fi MR }
\providecommand{\MRhref}[2]{%
  \href{http://www.ams.org/mathscinet-getitem?mr=#1}{#2}
}
\providecommand{\href}[2]{#2}

\clearpage

\setcounter{table}{0}
\renewcommand{\thetable}{S\arabic{table}}%
\setcounter{figure}{0}
\renewcommand{\thefigure}{SM\arabic{figure}}%
\setcounter{equation}{0}
\renewcommand{\theequation}{S\arabic{equation}}%
\setcounter{page}{1}
\renewcommand{\thepage}{SM-\arabic{page}}%
\setcounter{secnumdepth}{3}
\setcounter{section}{0}
\renewcommand{\thesection}{\arabic{section}}%
\setcounter{subsection}{0}
\renewcommand{\thesubsection}{\arabic{section}.\arabic{subsection}}%
\renewcommand{\thesection}{S-\Roman{section}}

\begin{center}
{\LARGE Supplemental Material}
\end{center}



\section{Linear response}
\setcounter{equation}{0}
\label{app:kubo}
We recall the derivation of the Kubo formula for the
conductivity tensor \cite{Kubo57}.
\subsection{Time evolution of the statistical operator}
\renewcommand{\CH}{H}
Consider a quantum system with Hamiltonian $\CH$ possessing a
discrete spectrum $(E_n)_{n=0}^\infty$ with corresponding eigenstates
$\{|n\>\}_{n=0}^\infty$.
Initially, at times $t < t_0$, the system is in equilibrium with
a heat bath in a state described by the statistical operator
\begin{equation}
     \r_0 = \frac1Z \sum_{n=0}^\infty \re^{- \frac{E_n}T} |n\>\<n|
\end{equation}
of the canonical ensemble. Here $T$ denotes the temperature and $Z$
the canonical partition function. At time $t_0$ a time-dependent
perturbation $V(t)$ is switched on adiabatically.

The time evolution operator $U(t)$ of the perturbed system satisfies
the Schr\"odinger equation
\begin{equation} \label{ut}
     \i \6_t U(t) = \bigl( \CH + V(t) \bigr) U(t)
\end{equation}
and the initial condition $U(t_0) = \id$. Under the influence of
the perturbation the eigenstate $|n\>$ evolves with time to
$|n, t\> = U(t) |n\>$, and the statistical operator at time $t$
becomes
\begin{equation} \label{rhot}
     \r (t) = U (t) \r_0 U^{-1} (t) \epp
\end{equation}
Setting
\begin{equation} \label{defW}
     W(s,t) = \re^{\i \CH s} V(t) \re^{- \i \CH s} \epp
\end{equation}
we see that, to linear order in the strength of $W$,
\begin{equation} \label{rhoborn}
     \r (t) = \r_0 - \i \int_{t_0}^t \rd t' \: [W(t' - t, t'), \r_0] \epp
\end{equation}
This is the statistical operator in Born approximation.
In the following $t_0$ will be sent to $ - \infty$ for
convenience.

\subsection{Time evolution of expectation values}
Using \eqref{rhoborn} we can calculate the time evolution of the
expectation value of an operator $B$ under the influence of the
perturbation. We shall denote the canonical ensemble average by
$\<\ \cdot\ \>_T = \tr \{ \r_0\ \cdot\ \}$ and write $B(t) = \re^{\i \CH t}
B \re^{- \i \CH t}$ for the Heisenberg time evolution of $B$.
Then
\begin{multline} \label{linresp}
     \de \< B\>_T (t)
        = \tr \{ (\r (t) - \r_0) B\}  \\
	= \i \int_{- \infty}^t \rd t' \: \tr \bigl\{ [\r_0, V(t')] B(t - t') \bigr\} \epp
\end{multline}
Here we have used the definition (\ref{defW}) of $W$, the cyclic
invariance of the trace and the fact that $H$ commutes with
$\r_0$. Equation (\ref{linresp}) is the `linear response formula'
\cite{Kubo57} which is utterly useful in many different contexts
in condensed matter physics.

Equation (\ref{linresp}) reveals more structure, if we assume
that
\begin{equation} \label{classpert}
     V(t) = - A F(t) \epc
\end{equation}
a classical force $F$ coupling to the system through an operator $A$.
Then
\begin{equation} \label{aborn}
     \de \< B\>_T (t) = - \i \int_{- \infty}^t \rd t' \:
        \tr \bigl\{ [\r_0, A] B(t - t') \bigr\} F(t') \epp
\end{equation}
If we define the response function
\begin{equation} \label{defrespfun}
     \PH_{BA} (t) = - \i \tr \bigl\{ [\r_0, A] B(t) \bigr\} \epc
\end{equation}
we see that the classical force $F$ and the response $\de \< B\>_T$
are connected through a linear integral operator with a kernel of
convolution type:
\begin{equation} \label{linrelt}
     \de \< B\>_T (t) =  \int_{- \infty}^t \rd t' \: \PH_{BA} (t - t') F(t') \epp
\end{equation}
Formally, this is a most general linear and causal relation between
an external force and the response of the system as seen in
the change of the expectation value of the operator $B$.

Now define the complex admittance
\begin{equation}
     \chi_{BA} (\om) = \int_0^\infty \rd t \: \re^{- \i \om t} \PH_{BA} (t)
\end{equation}
and take the Fourier transform of (\ref{linrelt}). Then the
convolution theorem implies that
\begin{equation} \label{genohm}
     \de B (\om) = \chi_{BA} (\om) F_F (\om) \epc
\end{equation}
where $\de B$ is the Fourier transform of $\de \<B\>_T$ and
$F_F$ is the Fourier transform of $F$.
\subsection{Alternative expressions for the response function}
An alternative formula for the response function can be
obtained noting that
\begin{multline}
     [\r_0, A] = \r_0 \bigl(A - \re^\frac{H}{T} A \re^{- \frac{H}{T}}\bigr)
	       = - \r_0 \bigl(A(- \i/T) - A\bigr) \\
	       = \i \r_0 \int_0^\frac{1}{T} \rd \la \: \dot A (- \i \la) \epp
\end{multline}
Inserting this expression into (\ref{defrespfun}) we obtain
\begin{equation}
     \PH_{BA} (t) = \int_0^\frac{1}{T} \rd \la \: \<\dot A (- \i \la) B(t)\>_T \epc
\end{equation}
implying the formula
\begin{equation} \label{scdadmitt}
     \chi_{BA} (\om) = \int_0^\infty \rd t \: \re^{- \i \om t}
                       \int_0^\frac{1}{T} \rd \la \: \<\dot A (- \i \la) B(t)\>_T
\end{equation}
for the complex admittance.
\subsection{Conductivity}
The force exerted onto a charged particle with charge $q$ by a
homogeneous electric field $\Ev (t)$ is $\Fv (t) = q \Ev (t)$, and
the corresponding potential energy at fixed $t$ is $- q \<\xv, \Ev(t)\>$.
Thus, a homogeneous electric field contributes
\begin{equation} \label{electricv}
     V(t) = - \sum_{j=1}^N \<\xv_j, \Fv(t)\>
\end{equation}
to the energy of a system of $N$ particles located at $\xv_j$,
$j = 1, \dots, N$.

$V(t)$ is a linear combination of perturbations of the form
(\ref{classpert}) considered above. Setting
\begin{equation}
     \Xv = \sum_{j=1}^N \xv_j
\end{equation}
we see that
\begin{equation} \label{defelectriccur}
     \dot \Xv = \Jv
\end{equation}
is the particle current operator.

Using (\ref{electricv})-(\ref{defelectriccur}) in (\ref{classpert})-%
(\ref{genohm}) and (\ref{scdadmitt}) with $\Av = \Xv$ and $\Bv = \Jv$,
we obtain the response of the current to a homogeneous force field,
\begin{equation} \label{ohmslaw}
     \de \Jv^\a (\om) = \Si^\a_\be (\om) \Fv_F^\be (\om) \epc
\end{equation}
where
\begin{equation} \label{conduct}
     \Si^\a_\be (\om) = \int_0^\infty \rd t \: \re^{- \i \om t}
                       \int_0^\frac{1}{T} \rd \la \: \<J_\be (- \i \la) J^\a (t)\>_T
\end{equation}
is the complex conductivity tensor.

This is the formula that is commonly used for the
calculation of the conductivity at finite temperature.
Alternatively, we may make direct use of the expression
(\ref{defrespfun}) and employ the cyclicity of the
trace in order to interpret it as a thermal average.
Inserting again $\Av = \Xv$ and $\Bv = \Jv$ we obtain
\begin{equation} \label{conduct2}
     \Si^\a_\be (\om) = - \i \int_0^\infty \rd t \: \re^{- \i \om t}
                                \<[X_\be, J^\a (t)]\>_T \epp
\end{equation}

\subsection{Response in non-thermal equilibrium}
The derivation of the linear response formulae (\ref{defrespfun})-%
(\ref{genohm}) is very robust. It relies on the smallness
of the perturbation $V(t)$ and on the fact that $\r_0$ is an
equilibrium distribution,
\begin{equation}
     [\r_0, H] = 0 \epp
\end{equation}
In particular, $\r_0$ need not be the statistical
operator of the canonical ensemble. Equations (\ref{defrespfun})-%
(\ref{genohm}) hold as well in the micro-canonical ensemble
or if $\r_0$ is the statistical operator of a generalized
Gibbs ensemble as it occurs in 1d integrable systems.

This implies, in turn, that (\ref{conduct2}) with the
thermal average replaced by an average over a generalized
Gibbs ensemble describes the conductivity within the linear
response regime for such systems.

Note, however, that (\ref{conduct2}) is no longer equivalent
to (\ref{conduct}) as the derivation of the latter formula
uses that the statistical operator involved in the ensemble
average is proportional to the time evolution operator for
imaginary times.

\section{The Drude model of conductivity}
For the interpretation of our results in the main text we
recall the Drude model \cite{AsMe76}. The Drude model is
a phenomenological model for the conductivity of solids based
on classical mechanics. A classical particle in a constant
homogeneous field would accelerate forever which is in
contradiction to Ohm's law saying that such a field induces
a constant current. In order to account for this fact Drude
introduced a friction term into the equation of motion of
an `average particle' at $\xv$ with `averaged velocity'
$\dot \xv$,
\begin{equation} \label{newton}
     m \, \ddot \xv (t) = \Fv (t) - \frac{m}{\tau} \dot \xv (t) \epp
\end{equation}
Here $m$ is the particle mass that we shall set equal to $m = 1/2$
as before; $\tau$ is interpreted as `collision time', an
average time between two collisions of particles, envisioned
as being reponsible for the friction term in (\ref{newton}).

Within this picture the average current per unit volume
is $\jv = \Jv/V = N \dot \xv/V = D \dot \xv$. Using
(\ref{newton}) we see that it satisfies the linear
differential equation
\begin{equation}
     \6_t \jv (t) + \jv (t)/\tau = 2 D \Fv(t) \epp
\end{equation}
Solving the latter by means of Fourier transformation we
end up with
\begin{equation}
     \jv_F (\om) = \s_\tau (\om) \Fv_F (\om) \epc
\end{equation}
where
\begin{equation}
     \s_\tau (\om) = \frac{2 \i D}{\om + \i/\tau}
\end{equation}
is the `Drude conductivity' determined by the single
phenomenological parameter $\tau$.

Note that
\begin{equation}
     \lim_{\tau \rightarrow + \infty} \Re \s_\tau (\om) = 2 \p D\, \de (\om)
\end{equation}
which agrees with the formula obtained for gases
within linear response theory.

\section{Kohn formula and TBA} \label{app:kohn}
In this supplement we use the Kohn formula for the calculation of the
finite-temperature Drude weight of the XXZ chain by use of Bethe ansatz and
TBA techniques and extensions as laid out in \cite{FuKa98,Zotos99}
(for a more recent example see \cite{LBPG25pp}). Here we use the formulation of
\cite{BFKS05}.

We start by writing equation (17) of the main text with abbreviating
the convolution integral of two functions by the symbol $\ast$,
\begin{equation} \label{yaya_mod}
     \log\eta = \beta\left(k^2 - \gamma k - \mu \right)
              - K \ast\log\left(1+\eta^{-1}\right) \epc
\end{equation}
where we also introduced $\eta$ related to $\varepsilon$ by
\begin{equation}
\eta=\exp(\varepsilon/T).
\end{equation}
The function $\varepsilon$ is known as the dressed energy 
and the function
\begin{equation}
\epsilon=k^2 - \gamma k - \mu \epc
\end{equation}
is known as the bare energy.
In TBA the density function $\rho$ of Bethe roots satisfies the linear
integral equation
\begin{equation}
(1+\eta)\cdot \rho=\frac 1{2\pi} p'-K\ast\rho\label{eqrho} \epc
\end{equation}
where $p$ is the bare momentum function and here of particularly simple form
$p(k)=k$ .

Central to the method is the calculation of two functions ${g}^{(1)}$ and
${g}^{(2)}$ describing the shifts of roots by finite size terms of order $1/L$
and $1/L^2$ in response to the modification of the periodic boundary condition
by a twist angle $\phi$. The functions ${g}^{(1)}$ and ${g}^{(2)}$ themselves
are of finite order and $\phi$ dependent. Derivatives with respect to $\phi$
will be denoted by dots.

Relations (9), (23), (25) and (26) of \cite{BFKS05} hold quite generally for many
integrable models. For the Bose gas many relations simplify, e.g. the
summations reduce to single terms. A trivial modification concerns the
arguments of the functions: here we use $k$ whereas $x$ is used in \cite{BFKS05}.

From \cite{BFKS05} we use (23)
\begin{equation}
(1+\eta)\cdot\left({\dot g}^{(1)} \rho\right)=\frac 1{2\pi}-K\ast\left({\dot g}^{(1)} \rho\right),
\label{eqrhodotg}
\end{equation}
Note that $p'(k)=1$, hence the two integral equations (\ref{eqrho}) and
(\ref{eqrhodotg}) are identical equations for the functions $\rho$ and
${\dot g}^{(1)} \rho$. As the solution is unique we find
\begin{equation}
  {\dot g}^{(1)}=1\label{eq3}
\end{equation}  
for all arguments.

Next we use (23) from which we take
\begin{multline}
{\ddot l}=\left(\rho+\rho^h\right)\left({\dot g}^{(1)}\right)^2+K\ast\left(\left({\dot g}^{(1)}\right)^2 \rho\right) \\ =\left(1+\eta\right)\rho+K\ast\rho=\frac 1{2\pi} \epc
\end{multline}
where in the second equation we used (\ref{eq3}) as well as $\eta=\rho^h/\rho$, and
in the last equation (\ref{eqrho}).

Then we take from (23) in \cite{BFKS05}
\begin{equation}
(1+\eta)\cdot\left({\ddot g}^{(2)} \rho\right)={\ddot l}'-K\ast\left({\ddot g}^{(2)} \rho\right)
\end{equation}
but here, the driving term ${\ddot l}'$ is just zero, so also the solution to
the integral equation ${\ddot g}^{(2)} \rho$ is zero, i.e.
\begin{equation}
  {\ddot g}^{(2)}=0 \epc
\end{equation}  
for all arguments.

In \cite{BFKS05} the Drude weight is defined differently
from here. Taking into account a factor $2\pi$ we
obtain from (25) in \cite{BFKS05}
\begin{multline}
  \Delta=\pi\int_{-\infty}^\infty\left(\epsilon''\left({\dot
        g}^{(1)}\right)^2\rho+\epsilon'  {\ddot
      g}^{(2)}\rho\right)dk \\ =2\pi\int_{-\infty}^\infty\rho dk=2\pi D \epc
\end{multline}
where we used the simplifications derived above as well as
the bare energy $\epsilon(k)=k^2 - \gamma k - \mu$ finally arriving at
the particle density $D$.

Note that (29) in \cite{BFKS05} does not hold for the Bose gas as the
bare energy is not identical to the derivative of the bare momentum (as is
true for the Heisenberg chain). Instead, for the Bose gas the derivative of
the bare energy is identical to the bare momentum whose derivative in turn is
the constant 1. We take this into account for deriving the analog of (29) in
\cite{BFKS05}. We obtain from (26) in \cite{BFKS05}
\begin{equation}
  \Delta=\frac\pi\beta\int_{-\infty}^\infty\frac{\rho\left(\frac\partial{\partial
        k}\log\eta\right)^2}{1+\eta^{-1}}dk \epc \label{interm}
\end{equation}
where we have directly used ${\dot g}^{(1)}=1$. Next we note an identity of
the derivatives
\begin{equation}
\frac\partial{\partial k}\log\eta=-2\frac\partial{\partial \gamma}\log\eta,
\end{equation}
taken at $\gamma=0$ which holds only in the thermal case. Inserting
this into (\ref{interm}) we obtain
\begin{multline}
  \Delta=\frac{4\pi}\beta\int_{-\infty}^\infty\frac{\rho\left(\frac\partial{\partial
        \gamma}\log\eta\right)^2}{1+\eta^{-1}}dk \\[-2ex] =
  4\pi\beta\int_{-\infty}^\infty\frac{\rho\left(\frac\partial{\partial
        \gamma}\varepsilon\right)^2}{1+\eta^{-1}}dk \epc \label{final}
\end{multline}
where finally we have used the relation between $\log\eta$ and the dressed energy
$\varepsilon$. Identity (\ref{final}) is nothing but (29) of the main text
upon inserting (28).


\end{document}